\documentclass[twocolumn,showpacs,preprintnumbers,amsmath,amssymb]{revtex4}

\usepackage{graphicx}
\usepackage{bm}

\begin{document}
\title{
Classical dynamics near the triple collision in a three-body Coulomb problem
}

\author{Nark Nyul Choi$^{\dagger}$}
\author{Min-Ho Lee$^{\dagger}$}
\author{Gregor Tanner$^*$}
\affiliation{$^{\dagger}$School of Natural Science, Kumoh National Institute
of Technology, Kumi, Kyungbook 730-701, Korea.}
\affiliation{$^*$School of Mathematical Sciences, 
University of Nottingham, University Park, Nottingham NG7 2RD, UK.}

\date{\today}

\begin{abstract}
We investigate the classical motion of three charged particles with both 
attractive and repulsive interaction.
The triple collision is a main source of chaos in such three body Coulomb
problems.
By employing the McGehee scaling technique, we analyse here for
the first time in detail the three-body dynamics near the triple collision
in 3 degrees of freedom.
We reveal surprisingly simple dynamical patterns in large parts of
the chaotic phase space.
The underlying degree of order in the form of approximate Markov partitions 
may help in understanding the global structures observed in quantum spectra 
of two-electron atoms.
\end{abstract}

\pacs{45.50.-j,05.45.-a,05.45.Mt,34.10.+x}
\maketitle

Our understanding of the overall dynamics of gravitational or Coulomb 
three-body problems (TBP) is still very limited. This can be attributed 
to the large dimensionality of the systems, the long range interactions 
and the complexity of the dynamics near the non-regularisable triple 
collision \cite{Dia97}. Contrary to gravitational TBP's, where the triple 
collision is of minor practical importance due to the overall stability of 
these systems, the dynamics in three-body Coulomb problems  
as for example two-electron atoms is dominated by the triple collision. 
Wannier's threshold law \cite{Wannier+Rost} as well as 
experimental results on double ionisation due to single \cite{Doe98} or 
multiple \cite{Web00} photon processes show electron-electron correlation 
effects which are linked to near triple collision events \cite{Ros03, Eck00}.  
The excitation of two single-electron wave packets in two electron 
atoms has recently been achieved experimentally making it possible to 
study classical collision events quantum mechanically \cite{Pish04}. The 
importance of the triple collision on quantum
spectra of two electron atoms is also evident from semiclassical treatments 
restricted to collinear subspaces of the full dynamics \cite{Tanner00}. 

We present here for the first time a comprehensive analysis of the 
classical dynamics near the triple collision in two electron atoms 
in the full $L=0$ phase space. Previous work has been restricted
to low dimensional invariant subspaces as for example collinear configurations 
or the so-called Wannier ridge \cite{Tanner00,Yuan+Sano,eZe+WR+Zee,WR,Zee}. 
We start by describing the dynamics {\em at} the triple collision in 3 
degrees of freedom. The topology of the space and the particle exchange 
symmetry strongly affect the scattering signal for 
energies below the three-particle break-up energy. We derive new scaling 
laws similar to the Wannier threshold law \cite{Wannier+Rost} and show for 
the first time that large parts of the phase space is highly structured 
in terms of approximate Markov - partitions. This is an important step  
towards analysing electron-electron correlation effects found 
in quantum properties of two-electron atoms \cite{Tanner00} in terms of
the overall phase space dynamics. 

Working in the infinite nucleus mass approximation and restricting
ourselves to angular momentum $L=0$, the dynamics may be 
described in terms of the electron-nucleus distances $r_1$ and $r_2$ 
and the inter-electronic angle $\theta$. Introducing the scaling 
parameter, $R = ({r_1^2 + r_2^2})^{1/2}$ and the hyper-angle
$\alpha = \tan^{-1} (r_2 / r_1 )$,
we employ the McGehee transformation \cite{McGehee,Yuan+Sano}
to obtain the equations of motion for two-electron systems
\begin{eqnarray} \label{EoM}
\dot{\alpha} = p_{\alpha}; \qquad \quad
&\quad \dot{p}_{\alpha}& = -\frac{1}{2}\, p_R\, p_{\alpha} -
     \frac{\partial}{\partial \alpha} {\bar{H}}; \\ \nonumber
\dot{\theta} =
     \frac{\partial}{\partial p_{\theta}} {\bar{H}}; \qquad
&\quad \dot{p}_{\theta}& =
- \frac{1}{2}\, p_R\, p_{\theta} -
\frac{\partial}{\partial \theta} {\bar{H}}; \\ \nonumber
\dot{\bar{H}} = p_R  \bar{H}; \qquad
& \quad \dot{p}_R &= \frac{1}{2}\, p_{\alpha}^2 +
\frac{1}{2} \frac{p_{\theta}^2}{\cos^2\alpha \sin^2\alpha} +
{\bar{H}}
\end{eqnarray}
where
\begin{eqnarray}\label{H_McGehee}
\bar{H} = \frac{1}{2} \left( p_R^2 + p_{\alpha}^2 +
\frac{p_{\theta}^2}{\cos^2\alpha \sin^2\alpha} \right) +
V(\alpha,\theta)= R E   , \\
V = -Z ( 1/\cos\alpha + 1/\sin\alpha )
    + 1/\sqrt{1- \sin 2\alpha \cos\theta}  , \nonumber
\end{eqnarray}
and $Z$ is the nucleus charge.
The scaled differential equations (\ref{EoM}) are independent of
$R$, and $R(t)$ and $\bar{H}(t)$ are given implicitly through
(\ref{H_McGehee}).
The triple collision itself has thus been removed from the dynamics and
the only singularities remaining are the two-body collisions at
$\alpha = 0, \pi /2$
which can be regularised by for
example considering the transformation $\bar{p}_{\alpha} =
p_{\alpha} \sin 2 \alpha$.

There are two fixed points of the dynamics (\ref{EoM}),
\[
   \alpha =\pi/4, ~\theta = \pi, ~p_{\alpha}= p_{\theta} = 0,
  ~ p_R = \pm P_0 ,
\]
where $P_0 = [{\sqrt{2}(4 Z -1)}]^{1/2}$.
These fixed points correspond to trajectories in the
unscaled phase space where both electrons fall into the nucleus
symmetrically along the collinear axis, that is, the
{\em triple collision point} (TCP) with $p_R = - P_0$ and
its time reversed partner, the trajectory of symmetric double
escape, the {\em double escape point} (DEP) with
$p_R = P_0$.  A true reduction in the dimensionality (from 5 to 4) 
of the problem is achieved for the special initial condition 
$\bar{H} = 0$ for which $\bar{H}$ becomes a constant of motion. 
The $\bar{H}=0$ - subspace which contains the two fixed points 
corresponds to $R = 0$ {\em or}
$E = 0$, that is, to the dynamics at the triple collision which 
is equivalent to the dynamics at $E=0$! We will come back to this
remarkable fact when studying the transition from $E=0$ to the 
dynamics for $E < 0$.  There are three invariant subspaces of
the dynamics: the collinear spaces $ \theta = \pi$, $p_{\theta} = 0$
(the $eZe$ configuration) and $\theta = 0$, $p_{\theta} = 0$
(the $Zee$ configuration) as well as the so-called
Wannier ridge (WR) of symmetric electron dynamics,
$\alpha = \pi/4$, $p_{\alpha} = 0$. The $eZe$ configuration
and the Wannier ridge are connected at the fixed points.

Linearising the dynamics near the DEP reveals that two of the five 
eigendirections, ${\bf v}^{(1, 2)}$, lie in the WR with eigenvalues
$\lambda_{D}^{(1, 2)} = - [1 \pm \sqrt{(4 Z -9)/(4 Z -1)} \, ] \, P_0 /4$;
the other three eigenvectors, ${\bf v}^{(3, 4, 5)}$, lie
in the $eZe$-space with
$\lambda_{D}^{(3, 4)} = -[1 \pm \sqrt{(100 Z -9)/(4 Z - 1)} \, ] \, P_0 /4$
and
$\lambda_D^{(5)} = P_0$, where the subscript $D$ refers to DEP
\cite{Wannier+Rost}. 
The unstable manifold $U_D^{(5)}$ along the ${\bf v}^{(5)}$
direction is oriented along $- p_R$ leading out of the $E = 0$
subspace; the unstable manifold $U_D^{(4)}$ lies in $E=0$
and corresponds to a trajectory leaving the DEP towards $p_R = \infty$.
The 3D stable manifold $S_D$
including $S_D^{(i)}$ associated with ${\bf v}^{(i)}$ ($i=1,2,3$)
is embedded in the 4D $E = 0$ subspace.
In what follows the structure of the $S_{D}$ is of special
importance being responsible for the striking features in the
classical electron - impact scattering signal for total
energy $E < 0$.
(From time reversal symmetry, one obtains for the TCP,
$\lambda_{T}^{(i)} = -\lambda_{D}^{(i)}$ for $i=1-5$.
The manifolds $U_{T}^{(1,2,3)}$ and $S_T^{(4,5)}$
can be obtained from the corresponding $S_D^{(1,2,3)}$ and $U_D^{(4,5)}$
respectively.)

\begin{figure}
\includegraphics[scale=0.48]{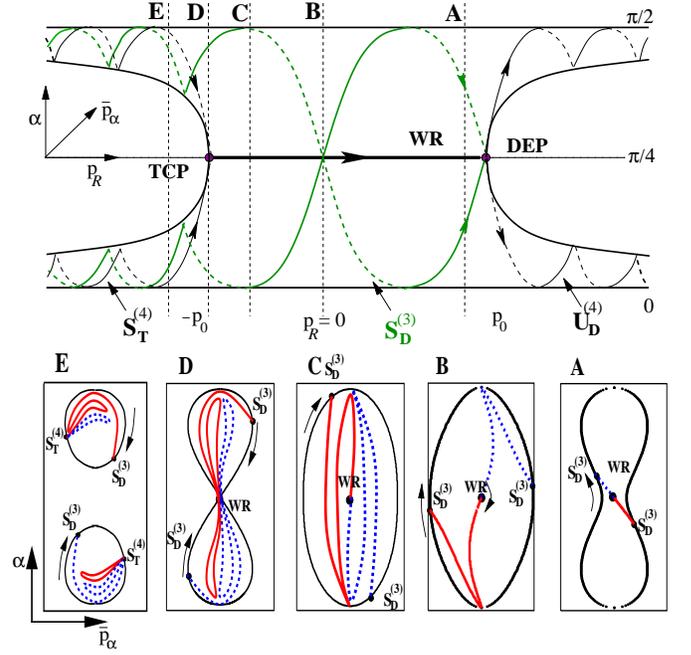}
\caption[]{\small  The PSOS $\theta = \pi$ of the 
$E=0$ manifold in $\alpha$-$\bar{p}_{\alpha}$-$p_R$
coordinates. The $eZe$ space forms the boundary
of the PSOS, the WR connects the TCP and DEP fixed points;
below, various cuts of the PSOS at fixed $p_R$ - values are shown 
together with the $S_D$. The two arms of the 
$S_D$ stretching from the WR towards the $S^{(3)}_D$ on the $eZe$ boundary 
are shown as full and dashed line, respectively. (The $S_D$ in {\bf C} and 
{\bf D} is drawn schematically to enhance important features.)
}
\label{fig:fig1}
\end{figure}

We analyse first the topology and the dynamics in the triple-collision
phase space $E= 0$ with a special focus on the structure of the $S_D$.
The {\em topology} of the space $E=0$ is most conveniently studied by
considering the 3D Poincar\'e surface of section (PSOS) $\theta = \pi$ in 
$\alpha$ - $p_{\alpha}$ - $p_R$ coordinates.
From (\ref{H_McGehee}),
one obtains that the $eZe$ - configuration forms the boundary
of the PSOS, see Fig.~\ref{fig:fig1}. A typical scattering event
follows a trajectory coming from infinity with $p_R = -\infty$,
$\alpha = 0$ or $\pi/2$ and one of the two electrons leaving
towards infinity with $p_R \to \infty$, $\alpha \to \pi/2$ or 0.
After switching to regularised coordinates $\bar{p}_{\alpha}$, the
topology of the $eZe$ - phase space takes on the form of a
sphere with four points taken to $|p_R| = \infty$, see
Fig.~\ref{fig:fig1} \cite{Yuan+Sano}. The two fixed points are
located at the saddles between the arms stretching in forward and backward
direction along the $p_R$ - axis. The WR space connects the TCP and DEP
along the axis $\alpha = \pi/4$, $\bar{p}_{\alpha} =0$, see 
Fig.~\ref{fig:fig1}; it is a compact manifold with the topology of
a $S^2$ - sphere where the fixed points form opposite poles.

For $\bar{H} = 0$, we have $\dot{p}_R \ge 0$, 
which leads to a relatively simple overall dynamics.
Its important features can be characterised
by the behaviour of the stable/unstable
manifolds of the fixed points. Various cuts of the $S_{D}$
in the PSOS for fixed $p_R < P_0$ are shown in
Fig.\ \ref{fig:fig1}. The $S_{D}$ is for $-P_0 \le p_R \le P_0$ bounded by
the 1D stable manifold $S^{(3)}_{D}$ in the $eZe$ - space,
and the 2D WR. That is, the $S_{D}$ is
the stable manifold of the WR. What makes the evolution of this manifold
remarkable, is its behaviour at the TCP at $p_R = -P_0$, where
the phase space itself splits into two distinct parts. Starting at
$p_R = P_0$ and going towards decreasing $p_R$ - values which
corresponds essentially to an evolution of the $S_{D}$ backwards
in time, the $S_{D}$ undergoes the usual stretching and folding
mechanism typical for an unstable manifold in compact domains.
The stretching and folding is here facilitated by an overall
rotation of the space around the WR - axis $\alpha = \pi/4,
\; \bar{p}_{\alpha} =  0$ and a certain ``stickiness''
near the binary collision points $ \alpha = 0$ or $\pi/2$, see the cuts 
{\bf B} and {\bf C} in Fig.\ \ref{fig:fig1} \cite{LATER}. 
As $p_R \to -P_0$, the phase space 
develops a bottleneck whereas the $S_{D}$ stretches over the whole phase space 
5 times by now. As $p_R$ decreases further passing through $-P_0$, the
$S_{D}$ is cut at the TCP into distinct parts, see {\bf D} in Fig.\ 
\ref{fig:fig1}. Points close to the TCP will leave the fixed point along 
the stable manifold $S_T^{(4)}$ in $eZe$ space towards $p_R \to -\infty$. 
In each arm exactly 5 pieces of the $S_{D}$ are connected to the 
$S_{T}^{(4)}$ for $p_R < - P_0$ forming two loops and 
one connection to the $eZe$ boundary. The $S_{T}^{(4)}$ itself is thus 
also a boundary of the $S_{D}$ without being part of it.
There are therefore two main routes towards the DEP for
electrons coming in from $p_R = -\infty$ close to the
$eZe$ - boundary: the first route approaches the TCP
on one of the 5 leaves of the $S_D$ close to the stable manifold
$S_{T}^{(4)}$ and moves then along the WR towards the DEP; secondly,
trajectories can approach the DEP 'directly' by moving on the
$S_{D}$ in the vicinity of the $S_{D}^{(3)}$. This twofold approach is
essential for understanding the dynamics in the $E<0$ - space.\\

We now turn to the full dynamics in the 5D $E<0$ phase 
space. In the following we analyse one-electron
scattering signals where electron 1, say, starts at
$r_1 = \infty$ with energy $E_1$ and fixed total energy
$E = -1$. We record the time delay
of the outgoing electron. (Note that fixing $E=-1$ is sufficient as
changing the total energy amounts to a simple scaling transformation
in the dynamics \cite{eZe+WR+Zee}). A smooth transition of the
dynamics from $E<0$ towards $E=0$ is achieved by
considering the limit $E_1 \to \infty$, $E_2 \to -\infty$ and
$E_1 + E_2 = -1$. The inner electron is then initially bounded
infinitely deep in the Coulomb well and interactions between the
incoming and bound electron will take place at $R\to 0$ and thus
$\bar{H} \to 0$. The dynamics in $E=0$ is in this sense
equivalent to the dynamics at the triple collision point $R =0$ and
the phase space flow at $E=0$ can be smoothly continued to the
flow in the full $E\ne 0$ space. Trajectories close to 
$\bar{H} =0$
will follow the dynamics in the triple collision space except
near the fixed points where the flow close to the manifold
$E=0$ is perpendicular to that manifold along the
${\bf v}^{(5)} = -p_R$ direction.
\begin{figure}
\includegraphics[scale=0.68]{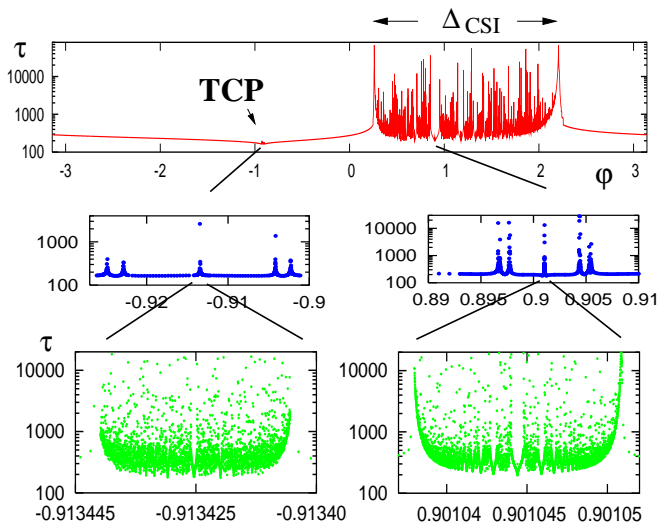}
\caption[]{\small The time-delay signal for $\theta = \pi/2$ and
 $E_1 = 0.2$; five distinct peaks appear in the 'dips' associated
with close encounters with the triple collision point. Note that delay
times are shown on logarithmic scales.
}
\label{fig:theta<pi}
\end{figure}

In Fig.~\ref{fig:theta<pi}, a typical signal for the 
scattering time $\tau$
is shown, here for $\theta = \pi/2$ and $E_1 = 0.2$, where
$\theta$ is the inter-electronic angle for $r_1\to\infty$. (We
have chosen zero angular momentum of both electrons initially).
$\varphi \in [-\pi, \pi]$ parametrises the starting point along one
revolution of the trajectory of the inner electron. Disregarding
the finer structures of the signal at the moment, two main features
emerge, a dip in the scattering time around $\varphi = -1.0$
and a chaotic scattering interval (CSI) around
$0.2 <\varphi<2.2$. The dip can be associated with trajectories close
to $S_T$ near $S_{T}^{(4)}$ which move 
towards the TCP. Most of these events will lead to immediate, fast ionisation
of one of the two electrons due to the large momentum transfer
possible near the triple collision singularity causing the dip in the 
scattering time.  Note that the 2D stable manifold $S_{T}$ is completely
embedded in the $eZe$ configuration; it is not possible to reach the
TCP from outside the $eZe$ region.

Chaotic scattering occurs, on the other hand, if trajectories
approach the DEP close to the stable manifold $S_{D}$ via the
direct route near $S_{D}^{(3)}$. The 3D stable
manifold of the DEP is part of the $E=0$ space and the DEP can thus
be reached for $E<0$ only in the limit $E_1 \to \infty$.
Orbits close to the $S_{D}^{(3)}$ will, however, also come close to
the DEP where they either follow $U_{D}^{(4)}$
leading to ionisation or follow $U_{D}^{(5)}$ along the
$p_R$ axis perpendicular to the $\bar{H} = 0$ - manifold. In the latter
case, $\dot{p}_R$ changes sign and electron trajectories fall back
towards the nucleus having equidistributed their momenta in such a way that
chaotic scattering becomes possible. The DEP thus acts as a turnstile for
entering into a {\em chaotic scattering region} of the full phase space. By 
time-reversal symmetry, the TCP acts as the exit gate for single electron 
ionisation.

A closer analysis of the strongly fluctuating signal in the CSI
reveals a series of dips flanked by singularities in the delay time on
either sides, see Fig.~\ref{fig:theta<pi}.
The dips correspond to orbits which after having spent some time in chaotic
motion leave the chaotic region by coming close to the TCP along the
stable manifold $S_{T}^{(5)}$. The borders of these intervals are
given by orbits escaping asymptotically with zero kinetic energy of the
outgoing electron.
All these escaping regions can be labelled uniquely by a finite
binary code reflecting the order in which binary near-collision events
take place after entering and before escaping the chaotic region.
The mechanisms leading to escape and to the existence of a binary symbolic
dynamics forming a complete Smale horse-shoe
are well documented and understood for the $eZe$ configuration
\cite{Yuan+Sano,eZe+WR+Zee}. It is surprising that
this mechanism holds in principle also far away from the $eZe$ regime
down to initial inter-electronic angles of the order
$\theta \approx \pi/4$.
 
There is one major difference between the scattering signal for
the collinear $eZe$ space and the dynamics for $\theta \ne \pi$; 
additional structures emerge at the centre
of the dips, see Fig.~\ref{fig:theta<pi}.
When enlarging the regions near the dips both at the primary dip and in
the CSI, one finds five separate peaks and, on further magnification, each 
of these peaks breaks up into a chaotic
scattering signal similar to the primary pattern.
One finds in this way a whole sequence of self-similar structures where
dips give birth to CSI's which in turn have 2nd generation
dips containing 5 peaks etc. Such a sequence leading up to the
second generation CSI is shown in Fig.~\ref{fig:theta<pi}.
\begin{figure}
\includegraphics[scale=0.85]{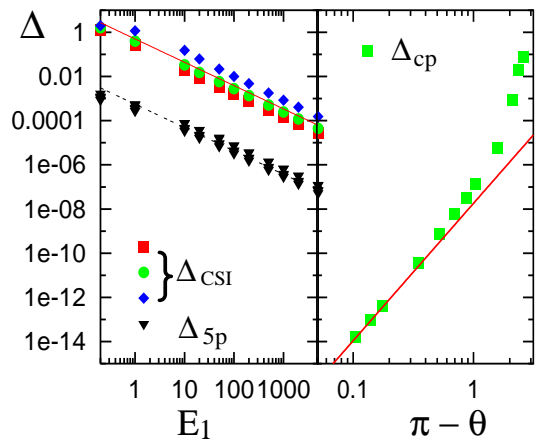}
\caption[]{\small Energy scaling of both the width of the
primary CSI $\Delta_{CSI}$ (see Fig.~\ref{fig:theta<pi})
for different $\theta$ values as well as the width of the 5 peaks 
$\Delta_{5p}$ (here for $\theta = \pi/2$) is shown; the scaling agrees 
with $\mu = 1.0558932 ...$ for $Z = 2$. To the right, the width, 
$\Delta_{CP}$, of the centre peak of the 5 peaks is shown as a function 
of $\pi - \theta$ (for fixed $E_1$) with scaling exponent 
$\nu = 6.223573 ...$.
}
\label{fig:DEP_scalall}
\end{figure}

These phenomena can be explained in terms of the dynamics in the
triple collision space $\bar{H} =0$ discussed earlier.
The dips represent close encounters with the TCP; whereas in the
$eZe$ - space, the only route away from the TCP is along the
unstable manifold $U_{T}^{(3)}$, for $\theta \ne \pi$ another route opens
up: escape along the WR and thus along parts of the $S_{D}$ leading
back to the DEP and to re-injection into the chaotic scattering region!
The existence of the 5 peaks in each of the dips is thus a clear manifestation 
of the stretching, folding and cutting mechanism of the $S_D$ in the 
$\bar{H}=0$ dynamics. A closer analysis reveals that the centre peak (CP) 
of the 5 peaks
corresponds to the part of the $S_{D}$ connected to the WR; the outer
peaks contain orbits which move away from the WR after passing the TCP
and before entering the DEP region. By repeatedly moving from the DEP
to the exit channel, the TCP, and then along the WR back to the
DEP, it is possible to create increasingly longer cycles of chaotic
scattering events. We do thus encounter here a rather curious dynamical feature,
namely a Smale-horseshoe, whose entrance and exit points are short-circuited
by a degenerate heteroclinic manifold, the WR, connecting the 
two fixed points sitting at these turnstile-gates.

The features described above imply scaling laws for the width $\Delta$ of the
CSI's for $E_1 \to \infty$. In this limit, trajectories which will enter the 
chaotic scattering region along $U_{D}^{(5)}$ need to come closer and closer 
to the DEP. The size of this phase space region near the DEP is limited by 
ejection along $U_{D}^{(4)}$. A simple estimate yields 
$ \Delta \sim E_1^{-\mu} $ with $ \mu = \lambda_{D}^{(4)}/\lambda_{D}^{(5)} $.
The {\em universality} of this scaling law for all CSI's is shown in
Fig.\ \ref{fig:DEP_scalall} and demonstrates that the DEP is indeed the 
sole entrance gate into the chaotic scattering region. The exponent $\mu$ is 
the so-called Wannier exponent which plays a crucial role in Wannier's 
threshold law for $E> 0$ \cite{Wannier+Rost} and is also of importance 
for quantum resonance widths \cite{Ros96}. A similar
scaling law can be deduced for the CP; the
phase space volume which can be transfered from the TCP to the DEP along
the WR scales in the $eZe$ - limit $\theta \to \pi$ like
$ \Delta_{CP} \sim (\pi - \theta )^{\nu}\, E_1^{-\mu} $
with
$ \nu = \lambda_{T}^{(3)}/ Re[\lambda_{T}^{(1)}]
    = \lambda_{D}^{(3)}/ Re[\lambda_{D}^{(1)}] $,
see Fig.\ \ref{fig:DEP_scalall}.

Our analysis suggests that the Markov partition leading to the symbolic 
dynamics of the $eZe$ - space remains largely intact over a substantial 
part of the $L =0$ phase space with the only modification that the 
number of symbols increases from 2 to $2 + 5 = 7$.
The existence of such a partition in this high-dimensional problem is
not obvious and may be the key for explaining 
approximate quantum numbers observed in two electron atoms in terms 
of the classical dynamics \cite{Tanner00}.

\begin{acknowledgments}
We thank the Royal Society (GT and NNC) and 
the Korea Research Foundation (KRF-2003-015-C00119) (NNC)
for finacial support,
and the Hewlett-Packard Laboratories in Bristol for their hospitality.
This work forms part of the EU-network {\em MASIE}. 
\end{acknowledgments}

\end{document}